\documentclass[10pt, conference, compsocconf]{IEEEtran}
\usepackage{ifpdf}
\usepackage{soul}
\usepackage{breqn}
\usepackage{rotating}
\usepackage[hyphens]{url}
\usepackage{color}
\usepackage[labelfont=bf,skip=0pt]{caption}
\usepackage{xspace}
\usepackage{chngcntr}
\usepackage{subcaption}
\usepackage[nodisplayskipstretch]{setspace}
\usepackage{diagbox}
\usepackage{graphicx}
\counterwithin{paragraph}{subsection}

\newcommand{\LONG}[1]{} 

\ifpdf
  \pdfoutput=1
\fi

\ifCLASSOPTIONcompsoc
  \usepackage[nocompress]{cite}
  \bstctlcite{IEEEexample:BSTcontrol}
\else
  \usepackage{cite}
\fi

\graphicspath{{images/}{graphs/}}
\DeclareGraphicsExtensions{.pdf}

\ifCLASSOPTIONcompsoc
  \usepackage[caption=false,font=footnotesize,labelfont=sf,textfont=sf]{subfig}
\else
  \usepackage[caption=false,font=footnotesize]{subfig}
\fi

\begin{document}
\title{Medusa: An Efficient Cloud Fault-Tolerant MapReduce}

\author{
\IEEEauthorblockA{Pedro~A.~R.~S.~Costa\IEEEauthorrefmark{1}, Xiao Bai\IEEEauthorrefmark{2}, Fernando~M.~V.~Ramos\IEEEauthorrefmark{1}, Miguel~Correia\IEEEauthorrefmark{3}\\
LaSIGE, Faculdade de Ci\^encias, Universidade de Lisboa, Portugal\IEEEauthorrefmark{1}, Yahoo Labs, Sunnyvale, USA\IEEEauthorrefmark{2},\\ INESC-ID, Instituto Superior T\'ecnico, Universidade de Lisboa, Portugal\IEEEauthorrefmark{3}\\
palcosta@fc.ul.pt, xbai@yahoo-inc.com, fvramos@ciencias.ulisboa.pt, miguel.p.correia@tecnico.ulisboa.pt}
}

\maketitle
\begin{abstract}
Applications such as web search and social networking have been moving from centralized to decentralized cloud architectures to improve their scalability.
MapReduce, a programming framework for processing large amounts of data using thousands of machines in a single cloud, also needs to be scaled out to multiple clouds to adapt to this evolution.
The challenge of building a multi-cloud distributed architecture is substantial.
Notwithstanding, the ability to deal with the new types of faults introduced by such setting, such as the outage of a whole datacenter or an arbitrary fault caused by a malicious cloud insider, increases the endeavor considerably.

In this paper we propose Medusa, a platform that allows MapReduce computations to scale out to multiple clouds and tolerate several types of faults.
Our solution fulfills four objectives.
First, it is transparent to the user, who writes her typical MapReduce application without modification.
Second, it does not require any modification to the widely used Hadoop framework.
Third, the proposed system goes well beyond the fault-tolerance offered by MapReduce to tolerate arbitrary faults, cloud outages, and even malicious faults caused by corrupt cloud insiders.
Fourth, it achieves this increased level of fault tolerance at reasonable cost.
We performed an extensive experimental evaluation in the ExoGENI testbed, demonstrating that our solution significantly reduces execution time when compared to traditional methods that achieve the same level of resilience.
\end{abstract}


\section{Introduction}
\label{sec:introduction}

MapReduce~\cite{Dean:10} has been attracting a lot of interest as a convenient tool for processing massive datasets using computer clusters.
MapReduce was developed and is widely used by Google, whereas one of its open source implementations, Apache Hadoop~\cite{White:09}, is currently used by several other companies, such as Yahoo and Facebook. 

Originally this framework targeted a single datacenter, so currently used implementations are not designed to work across multiple datacenters\footnote{\small We use the terms \emph{cloud}, \emph{cluster} and \emph{datacenter} interchangeably.}.
However, the last few years have witnessed the requirements for data- and compute-intensive analysis to grow significantly, increasing the need to scale-out computation across clouds.
As an indication of that direction, Baeza-Yates et al.~\cite{Baeza-Yates:2009:FMW:1645953.1646009} have assessed the feasibility of building distributed Web search engines comprising geographically dispersed sites.
The increasingly larger data sets used in bioinformatics have also led to the execution of multi-cloud computations for this type of applications~\cite{Matsunaga09cloudblast:combining}.
In addition, for a growing number of applications the data is gathered and stored in different datacenters -- i.e., it is inherently geo-distributed -- but the analyses required are global.
A well-known example includes the analysis of scientific data such as those originated from the Large Hadron Collider (LHC).
The tens of petabytes of data produced every year by this particle accelerator are stored in more than 140 datacenters distributed across 34 countries~\cite{Brumfiel2011}.

Acknowledging this trend, the research community has recently proposed MapReduce-based platforms that scale out to multiple clouds.
Wang et al.~proposed G-Hadoop~\cite{Wang2013739}, a new framework that enables large-scale distributed MapReduce computations distributed across multiple clusters.
The core of this system is the replacement of Hadoop's native file system, HDFS, with Gfarm, a distributed file system that scales to multiple clusters.
Another instance of this trend is G-MR~\cite{JayalathSE14}, another Hadoop-based framework that processes geo-distributed data across multiple datacenters.
These examples are demonstrative of the challenge to build multi-cloud distributed architectures.
Notwithstanding, the ability to deal with -- and to tolerate -- the new types of faults that are introduced by this setting makes building such system significantly harder.

Indeed, at scales of thousands of computers, switches, routers, power units and other components, failures are frequent.
Therefore, both the original MapReduce and Hadoop use two mechanisms to tolerate such faults: they monitor the execution of tasks and reinitialize them in case of failure (thus tolerating crash faults); and they add checksums to files that contain data to detect file corruptions~\cite{Ghemawat:03,White:09}.
Despite the use of such fault tolerance mechanisms, these MapReduce implementations do not consider three harder to tolerate failure modes.
First, some hardware faults may lead to the \emph{corruption of the processing}, leading to wrong outputs.
For instance, data may be corrupted while stored in DRAM or due to core chipset errors, two issues recently found to be reasonably frequent~\cite{Schroeder:09,Nightingale:11}.
This problem is, naturally, amplified when computations are scaled out to multiple datacenters.
Second, \emph{malicious attacks} perpetrated by corrupt cloud insiders (or by external hackers that attack a specific cloud) can also cause corruption of the processing.
The original MapReduce fault tolerance mechanisms cannot deal with such arbitrary~\cite{Avi04} or malicious faults~\cite{Ver03}.
Third, cloud outages may lead to the \emph{unavailability of MapReduce instances and their data}.
Experience shows that these events are also frequent, with cases of unavailability of hours to days in services like Windows Azure, Google Drive or Amazon EC2, to name just a few.
Again, current MapReduce systems cannot deal with cloud outage as they are restricted to work in a single datacenter.

To deal with such faults, one needs to add redundancy to the computation.
As both malicious faults and cloud outage can impair a complete cloud, handling them involves resorting to more than one cloud.
The terms \emph{cloud federation}~\cite{kurze2011cf} and \emph{cloud-of-clouds}~\cite{Bessani:2011:DDS:1966445.1966449} have been used recently to denote such virtual environments composed of multiple clouds.
We explore this idea to replicate MapReduce jobs in different clouds to avoid their incorrectness or unavailability due to the three kinds of faults explained before.

\LONG{  
There are several variables that we must consider when distributing a job between clouds in order to guarantee a good performance and correct execution.
In this work~\cite{BrunEBM11}, the authors developed a redundancy technique called \textit{iterative redundancy}, which ensures efficient replication of computation and data even when facing Byzantine faults.
This algorithm distributes a minimum number of redundant jobs across a node pool in order to achieve a desired confidence level in the result.
The number of the jobs to be distributed are defined based on the confidence level that the system must guarantee to the execution.
Differently, our algorithm measure the clouds based on the characteristics of the hosts and the network throughput to predict the execution time, and it is not suitable to work in a volunteer-computing system.
In the work of Verma~\cite{Verma:2011}, they predict the execution time by creating a job profile based on the counters that are in the logs from previous executions.
This algorithm calculates the number of tasks needed in order to finish an execution before a deadline.
This algorithm is made to work with a single Hadoop MapReduce runtime.
For us, the granularity of execution is focused on jobs running in several Hadoop MapReduce runtime in distinct clouds and not at the task level.
In addition, our proxy tolerates cloud outage which is something that the proposed solutions does not consider.
}

Tolerating cloud faults through replication is commonly considered expensive as one expects these faults to be rare.
The reality contradicts this expectation: cloud outages are becoming increasingly common~\cite{cloud-downtime-stats, Clarke}.
More importantly, for very critical applications such as bioinformatics or finance, any kind of errors and unavailability are unacceptable.
For example, a malicious insider in a cloud that hosts an epidemiological surveillance system that tampers with the diagnosis of patients may lead to disastrous consequences.
This particular problem is, today, a significant concern.
A recent report from the Cloud Security Alliance states malicious insiders as one of the top threats in cloud computing~\cite{alliance}, and alarming instances of this problem have recently occurred in companies such as Google~\cite{raey}.
Moreover, temporary unavailability of the financial system in one cloud (due to cloud outage) may dramatically influence the overall investment decision and cause huge financial loss.
Motivated by these scenarios, we believe the replication cost to be acceptable for such critical applications, in order to guarantee that rare faults with devastating consequences do not occur.
Interestingly,  cloud providers seem to share this concern: Amazon S3 recently launched  Cross-Region Replication to automatically replicate data across regions~\cite{AmazonWebServices2015}.

\subsection{Our proposal}

We propose a novel approach that allows MapReduce to scale out to multiple clouds to tolerate arbitrary and malicious faults, as well as cloud outages, for critical applications.
As per above, the use of multiple clouds for MapReduce is not in itself new.
The novelty of this work arises from the use of a multi-cloud environment to not only parallelize computation, but also to transparently tolerate different types of faults at the minimum cost.
Our solution addresses several non-trivial challenges for this purpose.
First, it aims to be a transparent solution for the user.
The MapReduce API is not changed, and the user simply writes her typical MapReduce application without modification.
Second, it does not require any modification to the Hadoop framework.
Third, it tolerates not only crash faults, as the original MapReduce, but also arbitrary faults, cloud outages, and malicious faults caused by corrupt cloud insiders.
Fourth, it achieves this level of fault tolerance at the minimum replication cost and guaranteeing acceptable performance.

Our approach relies on a \emph{proxy}, Medusa, that runs in the client and that interacts with (unchanged) MapReduce runtimes in different clouds to tolerate the three kinds of faults above.
The basic idea is to replicate each MapReduce job in more than one cloud and to compare the outputs of the replicated jobs to tolerate faults.
The challenge is to perform this efficiently, and doing so without changing the framework and keeping the whole process transparent to the user.
Achieving efficiency requires \textit{(i)} replicating each job the minimum number of times; and \textit{(ii)} assigning each replicated job to the cloud that ensures the best performance.
To this end, instead of replicating each job at least $2f + 1$ times to ensure a majority of correct results and tolerate $f$ faults, as is common~\cite{Sch90, Veronese:13}, our approach is crafted to run only $f+1$ replicas for each job when there is no fault, and $2f+1$ replicas when there are $f$ faults.

Importantly, our approach can tolerate not only $f$ but any number of faulty replicas or clouds as long as no more than $f$ faulty replicas return the same wrong output.
This includes the possibility that up to $f$ clouds maliciously collude and the system remains able to reach a correct output.
Moreover, we introduce a novel scheduling algorithm that takes into account the heterogeneity of the individual clouds to schedule the replicated jobs among them in order to reduce data communication and job completion time.

We performed an extensive experimental evaluation of our approach in a real testbed (ExoGENI).
The results demonstrate that our solution significantly reduces the execution time when compared to traditional methods that achieve the same level of resilience.
As an example, in certain scenarios we achieve a gain of up to 3 in efficiency when compared with a conventional round-robin approach that tolerates cloud faults.

In summary, our proposal is practical and transparent to the user by only involving a new software module running in the client (Medusa), not requiring any modification to the MapReduce framework or to user applications.
Simultaneously, it tolerates arbitrary and malicious faults, and cloud outages, efficiently.

\subsection{Outline}

The reminder of the paper is organized as follows. In the next section we describe the system model and define the problem. Then, in Section \ref{sec:solution:definition} we present the detailed design of Medusa, the cloud fault-tolerant MapReduce system we propose. Section \ref{sec:evaluation} reports on the experimental evaluation. In Section \ref{sec:relatedwork} we discuss related work and, finally, we conclude this paper in Section \ref{sec:conclusion}.

\section{Problem statement}
\label{sec:problem:statement}

To cope with the unprecedented data growth, applications such as web search, social networking and bioinformatic applications can be distributed in geographically-distant datacenters to leverage data locality and improve data processing efficiency~\cite{Baeza-Yates:2009:FMW:1645953.1646009, Matsunaga09cloudblast:combining}. In these cases, multiple MapReduce clusters are set up in distributed datacenters. Each cluster collects and processes the data that is close to its datacenter. The final data processing results are obtained by aggregating their respective outputs in an aggregation MapReduce job. In this work, we aim at providing cloud fault tolerance while minimizing the overall execution time for the MapReduce computations running in such multi-cloud systems.

\subsection{Hadoop MRv2}
\label{sec:hadoop:mapreduce}

MapReduce is a programming model for processing large-scale datasets with parallel and distributed algorithms in computer clusters. We focus in this work on Hadoop, which is an open source implementation of the original MapReduce. The Apache Hadoop platform includes the Hadoop kernel, Hadoop MapReduce, and Hadoop Distributed File System (HDFS).

\begin{figure}[!tb]
  \centering
  \includegraphics[trim=0mm -.5cm 4mm 0mm,width=.47\textwidth]{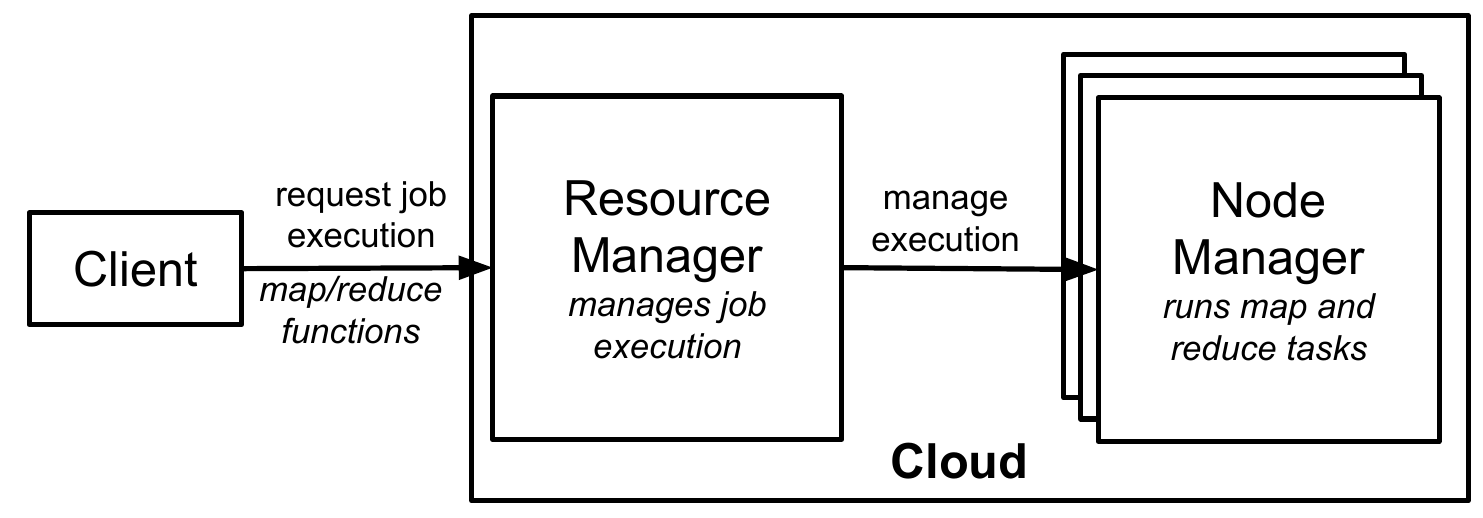}
  \caption{Simplified MapReduce execution process.}
  \vspace{1em}
  \label{fig:mr-architecture}
\end{figure}

Fig.~\ref{fig:mr-architecture} depicts the simplified execution process of MapReduce 2.0\footnote{\small Hadoop MapReduce has undergone a complete overhaul in Hadoop 0.23  and is now designated MRv2, or YARN.}. Specifically, a client requests the execution of a job from the \emph{resource manager} (previously called job tracker).
The role of the resource manager is to define which (map and reduce) tasks will be executed in which server.
The execution of a (map or reduce) task in a server is managed by the \emph{node manager} (previously called task tracker).

The Hadoop Distributed File System (HDFS) is the main distributed storage used by MapReduce applications. The input files for a MapReduce job typically reside in HDFS and are divided into logical blocks called splits. Each split is executed by a map task that produces an intermediate output in the form of key-value pairs. The content of the intermediate outputs is sorted by key and then consumed by the reduce tasks. The reduce tasks produce the final output.

Hadoop tolerates faults by \textit{(i)} monitoring and restarting tasks when servers, node managers or the tasks themselves crash through the resource manager; and \textit{(ii)} adding checksums to the files in HDFS to detect data corruption in disks. However, these mechanisms only work in a single cloud. Additional techniques are needed to tolerate faults for MapReduce jobs running in multi-cloud systems.

\subsection{System model}
\label{sec:problem:statement:system:model}

In a multi-cloud system, the MapReduce job runs in a federation of \emph{clouds}. Each cloud has an HDFS instance to store the initial inputs and final outputs of the jobs running in that cloud.
The entire data to be processed by the job is distributed across the clouds in the system, \textit{i.e.}, each cloud has a subset of the data stored in its HDFS. The data can be either collected by the cloud itself or assigned by some external processes, but we ignore this detail as it is application dependent and is orthogonal to the MapReduce execution.

\begin{figure}[!tb]
  \centering
  \includegraphics[trim=0mm -.5cm 3mm 0mm,width=.4\textwidth]{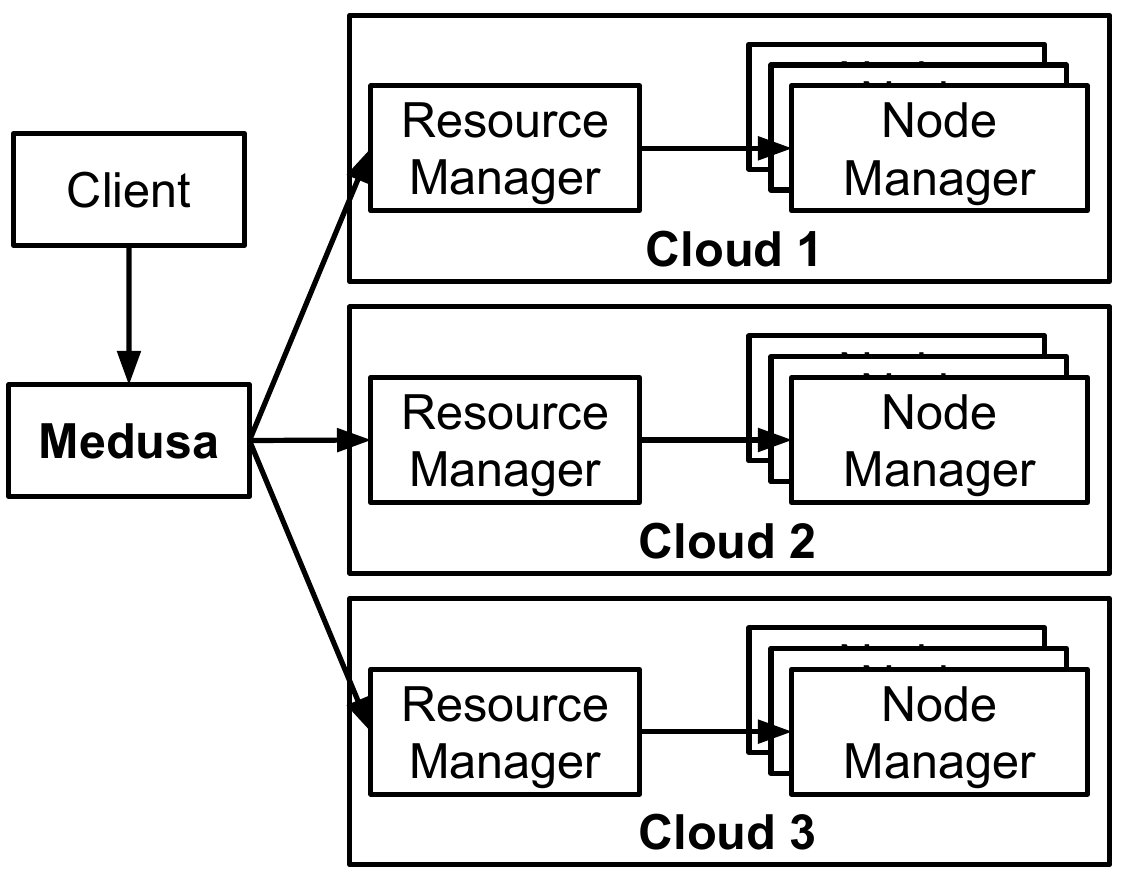}
  \caption{MapReduce in a multi-cloud system.}
  \label{fig:hadoop:manager:usecase}
  \vspace{1em}
\end{figure}

The MapReduce job is composed of a set of distributed \emph{processes} (Fig.~\ref{fig:hadoop:manager:usecase}). The \emph{client} submits the job through the process \emph{proxy}, Medusa, that controls the execution of the job in all the clouds. Each cloud first runs its own MapReduce instance to process the data it has: the resource manager controls the execution of the part of job assigned to that cloud; and the node manager in each server runs the map and reduce tasks assigned to that server. Finally, the proxy is in charge of assigning the outputs from these clouds to an aggregation MapReduce job in one of the clouds to obtain the final output for this job.

The messages between the proxy and the clouds are mediated by a message queuing service (MQ), which uses reliable channels so that no messages are lost, duplicated or corrupted.
In practice, this is provided by establishing TCP/IP connections.
A message is only lost if the cloud is unreachable.

The system is asynchronous, \textit{i.e.}, we make no assumptions about bounds on processing and communication delays in each stage of job execution (Section~\ref{subsec:solution:example}).
We assume the use of authenticated reliable channels for communication.

We assume the existence of one correct dataset in each cloud at the beginning of job execution.
We further assume the existence of a collision-resistant hash function, \textit{i.e.}, a hash function for which it is impossible to let two different inputs produce the same output (\textit{e.g.}, SHA-256).
We use the digests produced by this function to verify the integrity of the data replicated between clouds, and to validate the correctness of the job outputs.

The proxy is the key component in the multi-cloud system for tolerating cloud faults. We focus, in this work, on the design of the proxy as well as on a scheduling mechanism that guarantees good system performance.

\subsection{Fault model}
\label{sec:problem:faultmodel}

We say that a process is \emph{correct} if it follows the algorithm, otherwise we say that it is \emph{faulty}.
We assume that the client is correct, because it is the party interested in getting a correct job output.
We also assume that the proxy is correct, similarly to the Hadoop assumption that the resource manager does not fail.
The proxy is responsible for scheduling jobs among clouds to tolerate cloud faults and cloud outages.

Resource managers and node managers can fail arbitrarily: they can return wrong results (\textit{e.g.}, processing corruption) or even stop (\textit{e.g.}, cloud outage).
Malicious processes can also produce wrong results.
There is no limit on the number of faulty components of these kinds.
However, we assume that no more than $f$ resource managers -- \textit{i.e.}, $f$ clouds -- return to the proxy identical wrong outputs of the execution of a job.
This assumption allows the proxy to know that a result is correct by getting $f+1$ identical results.
The minimum number of clouds necessary to ensure termination is $2f+1$, as there may be $f$ faulty clouds.

\subsection{Problem formulation}
\label{sec:problem:statement:problem:definition}

We aim at tolerating \textit{(i)} arbitrary and malicious faults, and \textit{(ii)} cloud outages, when running MapReduce jobs in multi-cloud systems. To tolerate $f$ faults, a basic approach is to create $2f+1$ replicas of each job (\textit{i.e.}, the job running in each cloud and the aggregation job), spread them in $2f+1$ different clouds, and compare the $2f+1$ outputs of each job. If at least $f+1$ outputs are identical, their corresponding MapReduce jobs are correct and the identical output is the correct output of this job.

This basic approach has two major problems. First, it is expensive in terms of computation, communication, and storage. Even if there is no fault, each job is executed $2f+1$ times. This requires replicating the data initially hosted by each cloud to $f+1$ other clouds, which can be expensive in geographically distributed clouds. The same data also have to be stored in the HDFS of $2f+1$ clouds. In fact, if there is no fault, executing each job $f+1$ times is enough (there is no need for additional computations). Second, the basic approach does not take into account the difference among clouds for data replication and data processing. Intuitively, the data initially hosted by a cloud should be replicated to other clouds with which the original cloud has a high pair-wise bandwidth and, simultaneously, has high computational power. This would ensure the efficiency of the entire job execution in the multi-cloud system.

Therefore, our objective in this work is to design a MapReduce proxy that ensures the MapReduce job running in multiple distributed clouds to tolerate cloud faults while \textit{(i)} minimizing the amount of data replication and processing; and \textit{(ii)} ensuring efficient completion of the entire MapReduce job.

\section{Medusa: a cloud fault-tolerant MR}
\label{sec:solution:definition}

As mentioned in Section \ref{sec:problem:statement:system:model}, a full job execution in a multi-cloud system is comprised of two phases. The first phase runs a \emph{vanilla MapReduce job} in each cloud that holds a subset of data initially owned by that cloud. The second phase runs a \emph{global MapReduce job} that aggregates the outputs from all clouds to generate the final results. To tolerate arbitrary, malicious faults, and cloud outage, the MapReduce jobs in each phase need to be replicated to other clouds for ensuring the existence of $f+1$ identical outputs and thus the correctness of the results.

We propose in this paper a MapReduce proxy that works as a middleware in a multi-cloud system (\textit{i.e.}, a federation of clouds).
We refer to this proxy as \textit{Medusa}\footnote{Medusa is the mythological figure that has living snakes in place of hair. Metaphorically, the connections of our proxy to the clouds are the snakes, and these follow orders given by Medusa's brain (our proxy).}.

As explained, we assume that no more than $f$ replicas of a job return to the proxy identical wrong outputs from their executions. This assumption allows the proxy to know that a result is correct by getting $f+1$ identical results from different clouds / resource managers.
Given the expected low probability of arbitrary and malicious faults, it is too expensive to always execute $2f+1$ replicas of a job as is done in typical approaches.
Therefore, instead of replicating each job $2f+1$ times, Medusa first replicates each job $f+1$ times in
$f+1$ different clouds (\textit{i.e.}, once in the cloud where the data initially are and $f$ times in other clouds that do not have the corresponding data initially). If the executions of these replicas do not produce identical outputs, one more replicated job is launched in a different cloud until the proxy gets $f+1$ identical outputs. This \emph{deferred job execution} avoids the redundant data transmission, storage and processing when no fault happens.
As per the point above, each subset of data initially available in one cloud is replicated (at least) $f+1$ times, instead of $2f+1$ times.

We detail in Section \ref{subsec:solution:example} how Medusa works and explain in Section \ref{subsec:solution:scheduler} how the replicated jobs are scheduled among the clouds in the system to ensure efficient completion of the entire job.

\subsection{Medusa proxy in a nutshell}
\label{subsec:solution:example}

\begin{figure}
    \centering
    \includegraphics[trim=0mm -.5cm 3mm 0mm,width=.47\textwidth] {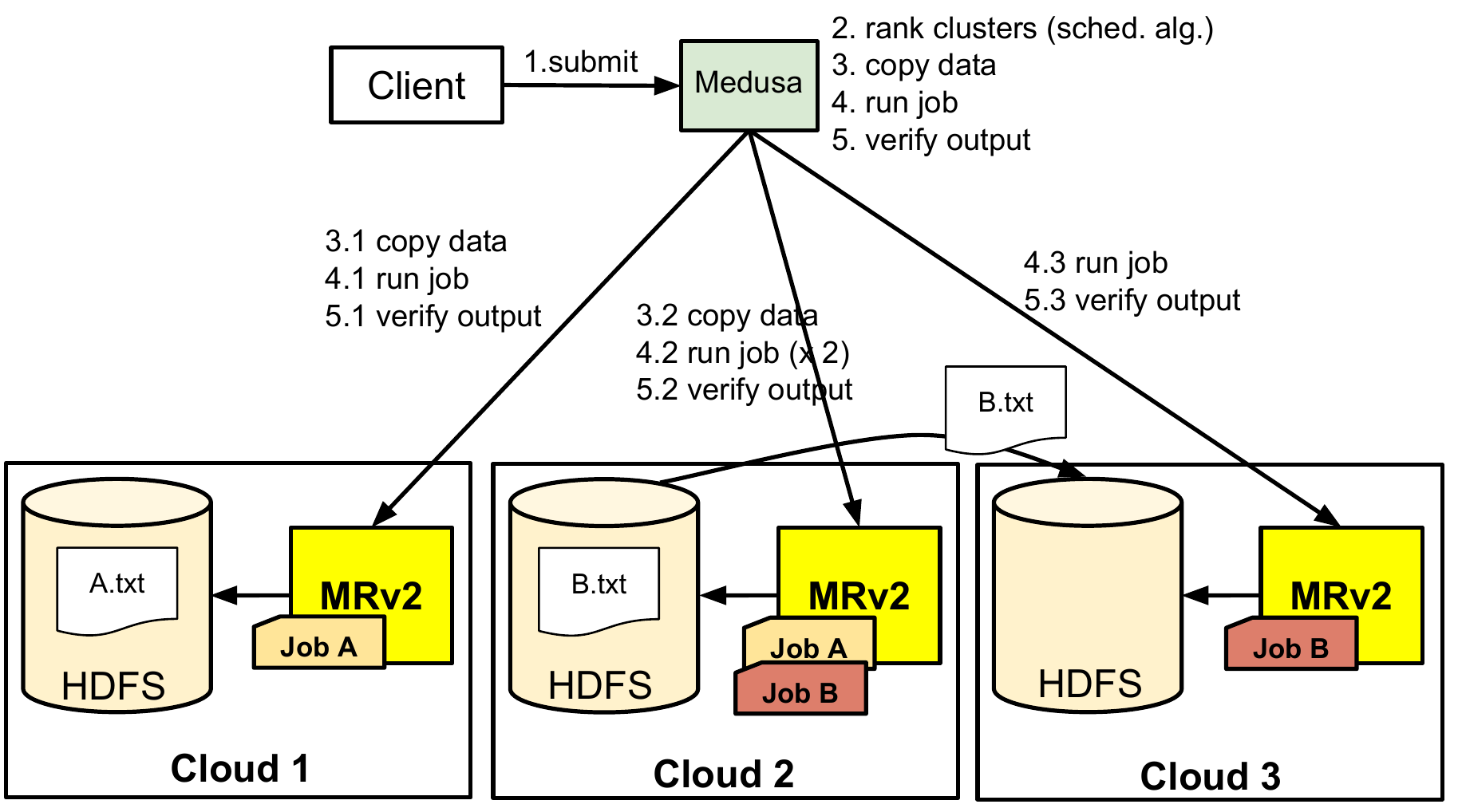}
    \caption{Phase 1: Vanilla MapReduce execution.}
    \label{fig:coc:runtime:diagram:phase:a}
    \vspace{.8em}
\end{figure}

\begin{figure}
    \centering
    \includegraphics[trim=0mm -.5cm 3mm 0mm,width=.47\textwidth] {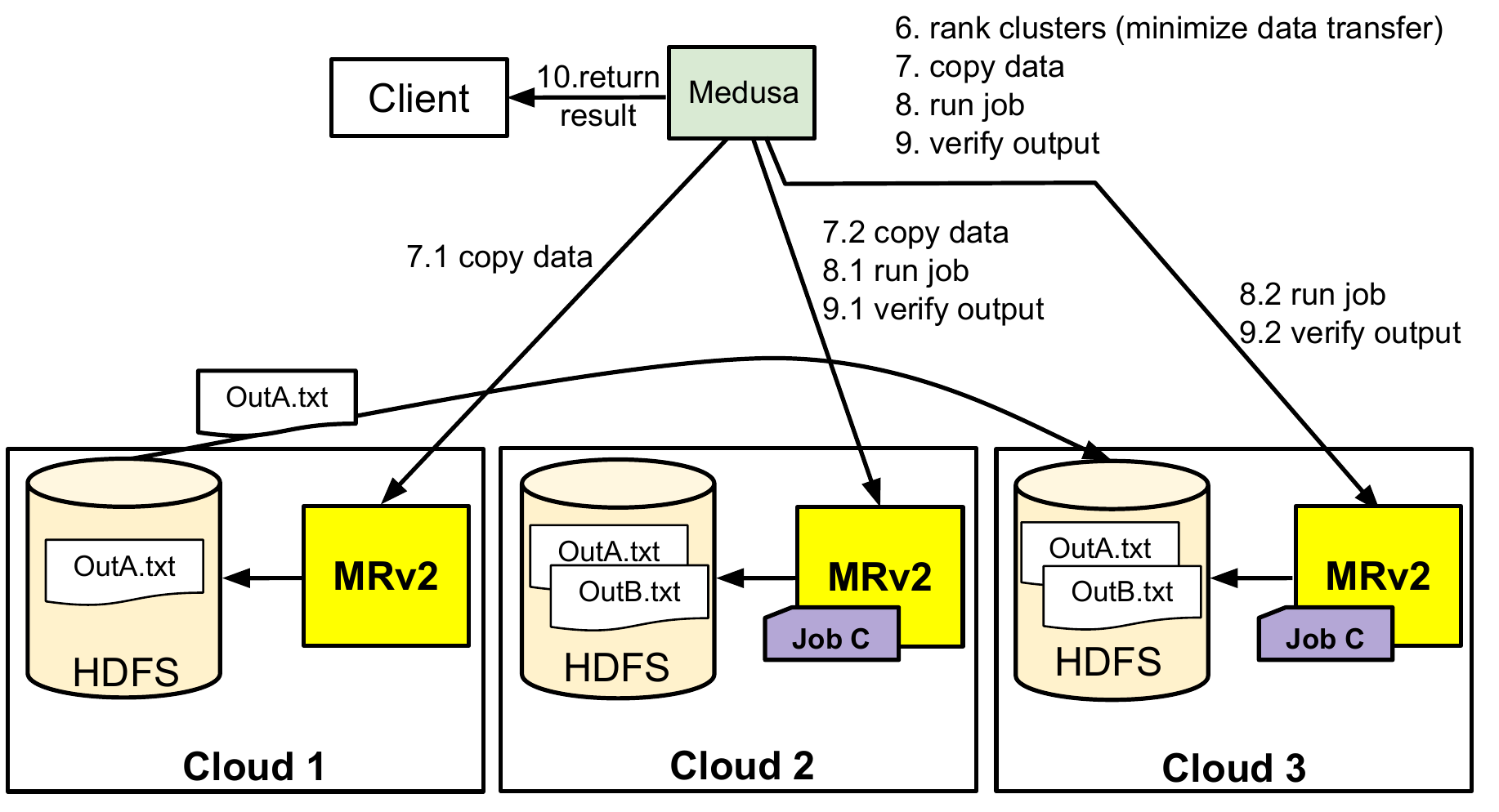}
    \caption{Phase 2: Global MapReduce execution.}
    \label{fig:coc:xruntime:diagram:phase:b}
    \vspace{1em}
\end{figure}

Each of the two phases of a full job execution is composed of three stages: replicating the data from the local cloud that initially holds them to other clouds; running the replicated jobs in all the clouds having the same data; and agreeing on the outputs of the replicated jobs. In each stage, all processes wait until every other process finishes, to move to the next stage. Fig.~\ref{fig:coc:runtime:diagram:phase:a} and Fig.~\ref{fig:coc:xruntime:diagram:phase:b} depicts the two-phase execution by means of an example, where the full job is running in 3 clouds and is set to tolerate 1 fault ($f=1$).

The proxy interacts with the clouds with the help of a message queuing service (MQ).
In Fig.~\ref{fig:coc:runtime:diagram:phase:a} and Fig.~\ref{fig:coc:xruntime:diagram:phase:b}, MQ is tightly coupled with the proxy, but in practice it can be running in a different host.

In the first phase (Fig.~\ref{fig:coc:runtime:diagram:phase:a}), the client submits a job (Step 1) with input data \emph{A.txt} and \emph{B.txt} (representing the two subsets of the entire data to be processed).
As mentioned earlier, we assume that the cloud collected or generated the data itself as we anticipate this to be the most common scenario.
Hence, the input data is already stored in \textit{cloud 1} and \textit{cloud 2}, respectively.
After receiving the request from the client, Medusa runs a scheduling algorithm (Section~\ref{subsec:solution:scheduler}) to select the best cloud(s) to run the replicated jobs (Step 2).
Medusa selects the clouds that offer the best performance in terms of data transmission and data processing to ensure low job makespan (\textit{i.e.}, the time it takes for the whole job to finish).
In this example, Medusa chooses to copy \emph{A.txt} to \emph{cloud 2} and \emph{B.txt} to \emph{cloud 3}.
Note that to guarantee data integrity, a digest of each data is computed using a collision-resistant hash function (\textit{e.g.}, SHA-256) and is sent with the original data for validation at reception.

If the data transmission is successful and the communication is not tampered, the execution of the job will start (Step 4).
After the executions of the $f+1$ job replicas finish, digests of their outputs are computed using the same collision-resistant hash function, and the correctness of these outputs are verified (Step 5).
If all the digests are identical, the vanilla MapReduce job finishes successfully.
Otherwise, the data is copied to a different cloud, selected by Medusa's scheduling algorithm, and the job is executed again to obtain $f+1$ identical outputs.
Once a majority of correct MapReduce jobs have finished, the second phase can start.

In the second phase (Fig.~\ref{fig:coc:xruntime:diagram:phase:b}), Medusa chooses the $f+1$ clouds that minimize the data transmission time to gather the outputs of the first phase and to run the global MapReduce jobs to produce the final output.
In this example, only \emph{OutA.txt} is replicated to \emph{cloud 3} (Step 7).
Then, after verifying the integrity of the data, the $f+1$ replicated jobs are launched (Step 8).
Once the $f+1$ replicated jobs finish, the digests of their outputs are computed and compared (Step 9).
As in the first phase, if the digests are identical, the job finishes successfully, and the correct result is returned to the client.
Otherwise, another replica of the global MapReduce job is launched until $f+1$ equal results from different clouds are obtained.

\subsection{The Medusa scheduler}
\label{subsec:solution:scheduler}

When a client submits a job, Medusa needs to instantiate $f+1$ replicas of the job.
These replicas will be launched in different clouds to tolerate cloud faults as explained in the previous section.
Whenever there is a disagreement on the output of a job, Medusa needs to launch an extra execution of that job.

Deciding in which cloud a replicated job should be executed is crucial for the performance of the system. Specifically, the client is interested in minimizing the time it takes for the submitted job to complete: the \emph{makespan} of the job. Intuitively, if a job, as well as the data it needs to process, is replicated to a particular cloud with high computational power and to which it is connected by high-bandwidth links, it should take relatively shorter time for the job to complete, when compared to the available alternatives. In contrary, if the data has to be copied to a cloud using low bandwidth links, or to a cloud that is already overloaded, it may take a very long time for the job to complete. Therefore, to ensure a small makespan of the entire job, it is important for Medusa to choose the appropriate clouds to replicate each job.

To this end, we propose a scheduler to distributes the replicated jobs across different clouds based on the predicted data transmission time and data processing time in each cloud. The prediction takes into account the historical performance as well as the current status of each cloud, allowing us to incorporate the \emph{heterogeneity} of the clouds into the scheduling decision.

The overall makespan is determined by the longest time to complete the vanilla MapReduce jobs running in all the clouds and the time to complete the global MapReduce job.
Each job is replicated and executed between $f+1$ times and $2f+1$ times, depending on the number of faults $f$. The Medusa scheduler follows a greedy approach and selects, for each job, the clouds that minimize the time to complete the executions of all its replicas. Specifically, the Medusa scheduler estimates, for each job, the time to replicate and process the corresponding data in each cloud that does not initially holds the data, and selects the $f+1$ clouds in increasing order of the estimated time. In case of a fault, the Medusa scheduler makes another estimation and chooses the cloud with the shortest estimated time to run the extra replica.
Note that the scheduler will deal in accordance with the type of fault.
When the system is set (by the client) to assume any fault can be malicious, it will not consider the cloud where the fault occurred as an option to run the extra replica.
That cloud cannot be trusted again since the malicious insider may corrupt the results once more.
When the system is set to only tolerate arbitrary faults and cloud outage, any alive cloud can be an option to launch the extra replica.

Formally, the estimated time $t_1(i)$ for completing a (replicated) vanilla MapReduce job (\textit{i.e.}, phase 1) in cloud $i$ can be written as
\begin{equation}
    \label{eq:time:eq1}
    t_1(i) = t_{trans}(j, i) + t_{1, proc}(i),
\end{equation}
where $t_{trans}(j, i)$ is the estimated time to transfer the data to be processed by the job from cloud $j$ to cloud $i$, and $t_{1, proc}(i)$ is the estimated time to execute the vanilla MapReduce job in cloud $i$.

The estimated time $t_2(i)$ for completing a (replicated) global MapReduce job (\textit{i.e.}, phase 2) in cloud $i$ can be written as
\begin{equation}
    \label{eq:time:eq2}
    t_2(i) = max_{j\in C\wedge j\neq i}\{t_{trans}(j, i)\} + t_{2, proc}(i).
\end{equation}

This equation requires further explanation. For the global MapReduce job, all the outputs produced by the vanilla MapReduce jobs need to be copied to cloud $i$ except the ones that already exist in cloud $i$. As the data transmission can take place in parallel, the time for transmitting the outputs of the vanilla MapReduce jobs is bounded by the maximum time for transmitting the outputs from any cloud $j$ to cloud $i$ ($t_{trans}(j, i)$). $t_{2, proc}(i)$ is the estimated time to execute the global MapReduce job in cloud $i$.

Therefore, for each job the Medusa scheduler selects the cloud $i$ that has the shortest estimated time $t_1(i)$ or $t_2(i)$ to run a replica among the clouds that have not been selected until the correct output is obtained. We explain in the following how the data transmission time $t_{trans}(j, i)$ and the data processing time $t_{1, proc}(i)$ and $t_{2, proc}(i)$ are estimated, respectively.

\paragraph{Estimating data transmission time} The data transmission time between two clouds depends on \textit{(i)} the network distance between them, \textit{(ii)} the network throughput between them, and \textit{(iii)} the size of the data to transfer. Hence, we estimate the time to transfer data of size $S$ between cloud $j$ and cloud $i$ as
\begin{equation}
  \label{eq:time:eq3}
  t_{trans}(j, i) = l(j, i)/2 + S/T(j, i),
\end{equation}
where $l(j, i)$ is the round-trip time between cloud $j$ and $i$, and $T(j, i)$ is the estimated throughput between them.

The value of $l(j, i)$ depends on the geographical distance between clouds $j$ and $i$, the speed of transmission in the different propagation media connecting them, among other variables, and is usually stable. Since this value is small (typically in the milliseconds range), accurate estimation of the data transmission time depends largely on the estimation of the network throughput $T(j, i)$ between cloud $j$ and cloud $i$.
Nevertheless, we decide to include both variables in Equation~\ref{eq:time:eq3} for the sake of model completeness.

Considering that the throughput varies depending on the traffic load of other connections, Medusa keeps track of the throughput for each pair of clouds in the system.
The measures are taken sequentially and periodically by running the network tool \emph{Iperf}, a tool that measures maximum TCP bandwidth.
Specifically, a script is running in each cloud to measure its throughput to other clouds.
Before scheduling a job, Medusa will access this information to estimate the throughput from one cloud to the others. The throughput between two clouds is estimated as the average throughput between them over a window of size $k$, where $k$ is the number of the most recent throughput measurements.

\paragraph{Estimating data processing time} The time for completing a given MapReduce job mainly depends on \textit{(i)} the capacity of the cloud running this job and \textit{(ii)} the configuration of the job. Obviously, if a cloud has high computational power and large amounts of free resources, the job running on it will finish within a short time. In addition, a high level of parallelization (\textit{i.e.}, large number of map and reduce tasks) for the same job in the same cloud implies shorter data processing times. Considering this, the Medusa scheduler relies on a linear regression model to predict the data processing time for a MapReduce job to complete in a cloud.

Specifically, the Medusa scheduler trains one linear regression model for each cloud in the system to predict its time to complete a MapReduce job, in the form
\begin{equation}
  \label{eq:linear-regression:def:eq}
  \hat{y} = \beta_1x_1 + \dots + \beta_nx_n + \beta_0, \\
\end{equation}
where $\hat{y}$ is the data processing time to predict (\textit{i.e.}, $t_{1, proc}(i)$ or $t_{2, proc}(i)$ in cloud $i$) and $x_1$, ..., $x_n$ are the $n$ features we use to make the prediction.
We estimate the parameters $\beta_0$, ..., $\beta_n$ using the least squares approach~\cite{bjoerck_least_squares96}.

The Medusa scheduler relies on three types of features to make the prediction: \textit{(i)} job configuration; \textit{(ii)} cloud capacity; and \textit{(iii)} cloud overhead. We describe the representative features for each type in the following.
\begin{itemize}
    \item \textit{Job configuration features.} We consider the size of the input data, the number of map tasks and the number of reduce tasks as features in this type. Clearly, large input data and a small number of map and reduce tasks imply long job completion times. The values of these features are always known to the Medusa scheduler.
    \item \textit{Cloud capacity features.} We consider the clock speed (MHz) and the number of cores of the CPU and the total memory capacity (MBs) as features in this type.
    These are variables that define the capacity of cloud, but they do not tell the load of the cloud in a specific time.
    \item \textit{Cloud overhead features.} In addition to the computational capacity, the overhead in the cloud also has an impact on the completion time of the job to schedule. For instance, if a cloud is overloaded and there are already a number of jobs queued to be launched, the scheduled job will not finish in a short time even if very small. In contrast, if a cloud has more free resources the scheduled job can finish early even if its capacity is relatively low. We use the number of MapReduce jobs that are currently running in the cloud, the percentage of completion of the running MapReduce jobs, the number of MapReduce jobs that are queued to run, and the size of the input data of the running jobs as features in this type. These are part of the filesystem information that MRv2 can provide.
\end{itemize}

For each vanilla MapReduce job to schedule, the Medusa scheduler estimates the data processing time $t_{1, proc}(i)$ in cloud $i$ using this linear regression model.
For the global MapReduce job, since the data to be processed is typically orders of magnitude smaller than the original data (thanks to the vanilla MapReduce jobs), it can be efficiently executed in any cloud.
We thus ignore the time for running the global MapReduce job (\textit{i.e.}, $t_{2, proc}(i)$) when making the estimation of job completion time (\textit{i.e.}, $t_2(i)$) for cloud $i$. This helps to improve the efficiency of the Medusa scheduler.

\section{Evaluation}
\label{sec:evaluation}

This section evaluates the performance of our system.
Section \ref{subset:setup} describes the experimental setup as well as the implementation and configuration of Medusa.
Section \ref{subset:perf} reports on its performance, considering both the presence and absence of faults during job execution.

\subsection{Experimental setup}
\label{subset:setup}

\paragraph{Applications} 
We evaluate the performance of Medusa using the \textit{WordCount}, \textit{WebdataScan}, and \textit{Monsterquery} benchmarks from Hadoop's Gridmix benchmark \cite{gridmix}, as examples of applications commonly used in real-world scenarios.

Running \textit{WordCount} in a multi-cloud system can be considered as building the inverted indexes of a multi-site web search engine for each search site (\textit{i.e.}, cloud).
Specifically, in a multi-cloud web search engine each site is in charge of collecting and indexing a subset of the entire document collection. To build the search index (that supports the term frequency-inverse document frequency style ranking functions, TF-IDF), each search site runs a local MapReduce job to parse the documents it has and to count the occurrences of each term in a document (\textit{i.e.}, TF) and the number of documents in that search site containing each term (\textit{i.e.}, partial IDF). This can be achieved by running WordCount as a vanilla MapReduce job in each cloud. To ensure the same search results can be retrieved as in a single-cloud search engine, the local outputs from previous executions need to be aggregated to obtain the number of documents in the search engine that contain each term (\textit{i.e.}, global IDF). To this end, we implemented a \textit{WordCountAggregator}. This corresponds to the global MapReduce job described in Section \ref{sec:solution:definition}.
\textit{WebdataScan} is a benchmarking application that extracts samples from a large data set, which is a common form of processing in many systems.
\textit{Monsterquery} is another benchmarking application that queries part of the data from a large data set.
The MapReduce framework divides a query into steps and the dataset into chunks, and then runs those step/chunk pairs in separate physical hosts.
The mappers  perform the data collection phase and the reducers take care of data processing.

For evaluating these applications we used up to 6GB of data generated by Gridmix, equally partitioned and stored in all clouds.
For a small subset of the experiments we tested larger files.
The results obtained using these larger datasets confirm the general trend we report next.

\paragraph{Implementation and configuration}
We evaluated the system in the ExoGENI testbed\footnote{\url{http://www.exogeni.net}}, a distributed networked infrastructure-as-a-service spread across the USA that allows setting up virtual topologies across sites and servers in each site.

We set up four clouds located in different sites for each experiment.
For the WordCount benchmark, we used clouds  located in the East and West coasts of the USA: California, Chicago, and West-Virginia.
In the WebdataScan and Monsterquery benchmarks the clouds were geographically closer: Pittsburgh, Massachusetts, and Texas.
For the sake of heterogeneity we set for each experiment two clouds in the same state, with the other two in different states (\textit{e.g.}, in the WordCount experiment we set two clouds in Chicago).
The hardware used in all applications was diversified in terms of CPU characteristics and RAM size.
We have set one specific cloud in each experiment with better resources than the remaining ones in order to maximize heterogeneity and to demonstrate the benefits of the Medusa scheduler.
In the WordCount experiments the best cloud was in Chicago, whereas in the WebdataScan case it was in Pittsburgh.

Each cloud is composed of 4 hosts with a MRv2 runtime: one \emph{resource manager} (master) and 3 \emph{node managers}  (slaves).
The MRv2 framework is not modified, which leads to each Resource Manager being a single point of failure \emph{in its cloud}.
However, Medusa has the capability to detect if the MRv2 in a cloud is running.
If not, that cloud is considered faulty (a type of fault our system tolerates).
Medusa is installed in the client machine because we assume that the client is always correct. If a client was faulty, the job output and the proxy results would be compromised.


Medusa is implemented in Python 2.7. 
Medusa is the key component to schedule the jobs and tolerate cloud faults. It submits and coordinates the execution of jobs, makes scheduling decisions, and verifies the integrity of the replicated data and job outputs (checking if $f+1$ replicas are identical and launching new replicas accordingly).
These operations require the proxy to communicate with the Resource Manager in each cloud. The proxy is logically located outside the system (in the sense that the clouds are oblivious to it). In our experiments, it  runs in the same machine as the client, an Amazon AWS host located in Oregon.
The message queuing service used is RabbitMQ~\cite{videla2012rabbitmq}. The RabbitMQ server runs in the same machine as the proxy.

Each experiment is configured to first execute the WordCount, WebdataScan, or Monsterquery job and then aggregate the results to obtain the final output.
Each experiment is repeated 40 times. The Medusa scheduler relies on a history of 30 executions to train the linear regression model and estimate the throughput between clouds.
We use Hadoop's DistCp3 tool\footnote{Distributed Copy, or DistCp, is a tool that uses MapReduce for large inter/intra-cluster data copying.} to copy data between the HDFS of two clouds.
We inject random workloads by running random numbers of extra jobs in each cloud, to simulate background overhead added by other users.

\paragraph{Baseline} We compare the performance of the Medusa scheduler in terms of job makespan -- the time it takes to complete the entire job -- and system workload against a baseline that also tolerates cloud faults but uses a simple strategy to schedule the replicated jobs -- a \textit{Round-robin} scheduler.
This scheduler selects the clouds to run the replicas of a job in a circular order, assuming the clouds are numbered sequentially.


\subsection{Experimental performance}
\label{subset:perf}

\begin{figure*}[t!]

  \begin{subfigure}{0.31\textwidth}
    \includegraphics[trim=1mm 0mm .6cm 0mm,width=\linewidth]{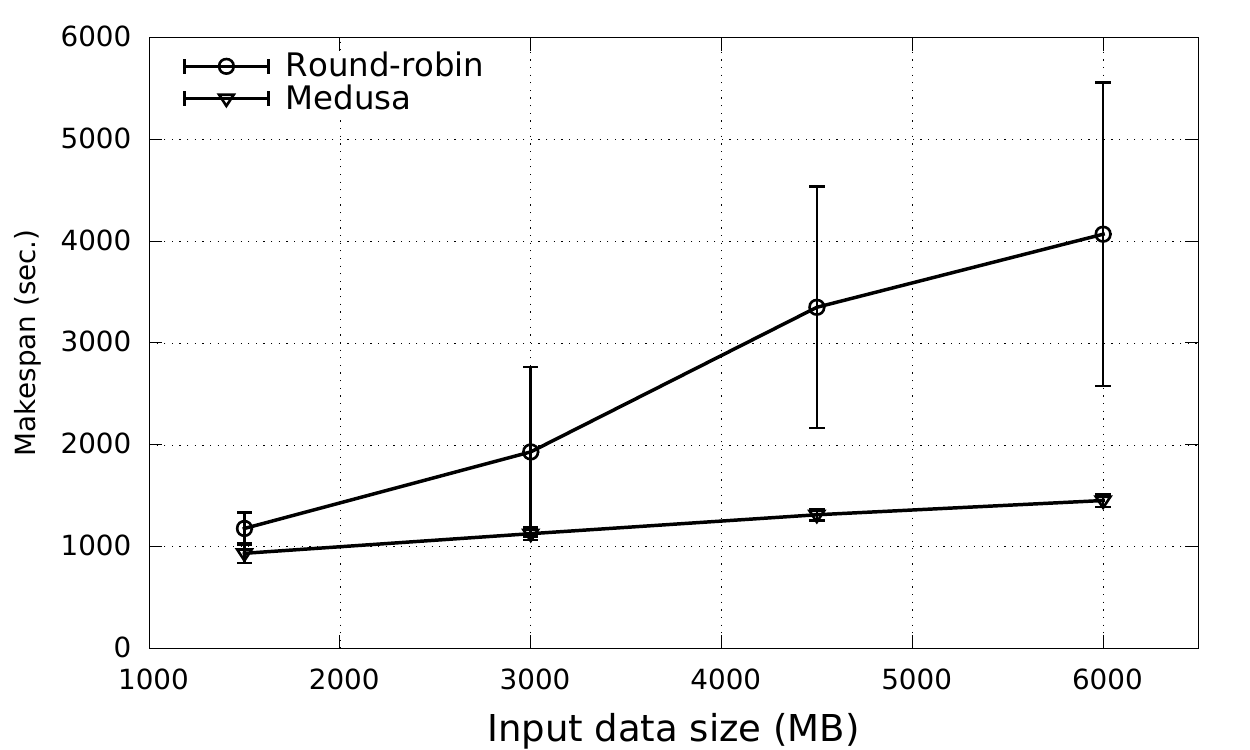}
    \caption{WordCount}
    \label{fig:makespan:performance:trendline:wordcount}
  \end{subfigure}
  \hspace*{\fill} 
  \begin{subfigure}{0.31\textwidth}
    \includegraphics[trim=1mm 0mm .5cm 0mm,width=\linewidth]{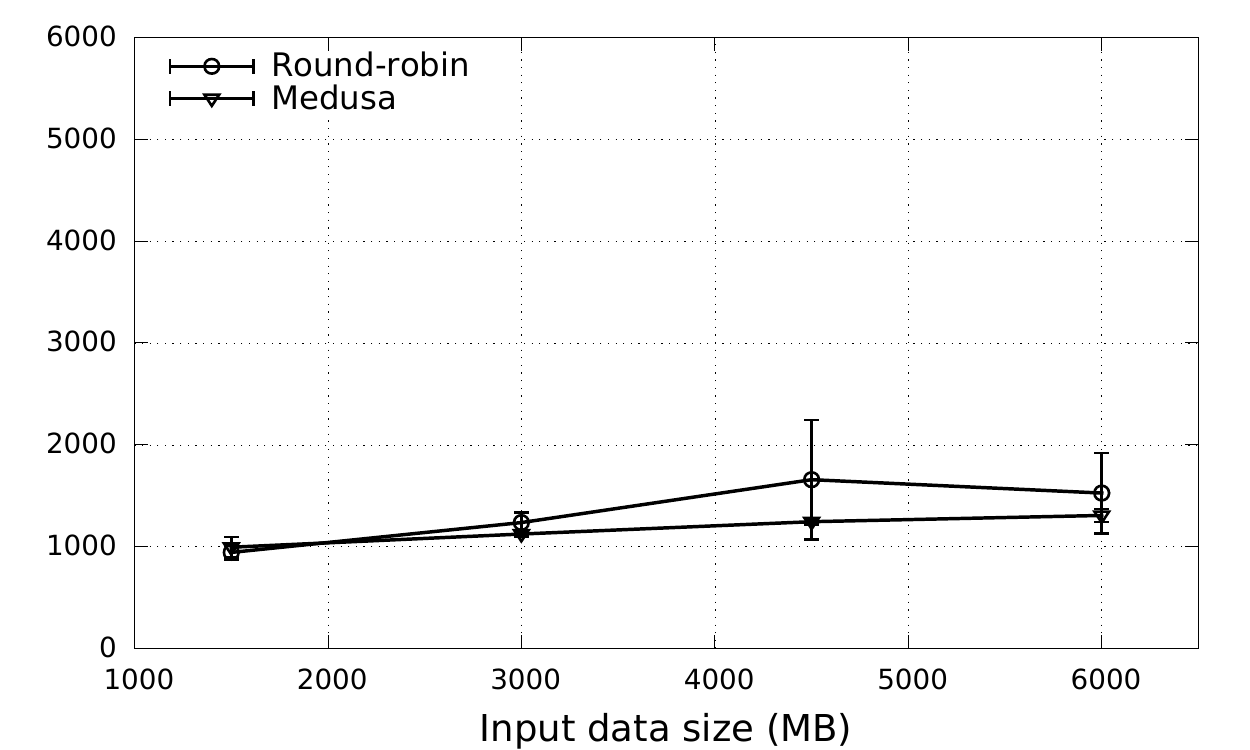}
    \caption{WebdataScan}
    \label{fig:makespan:performance:trendline:webdatascan}
  \end{subfigure}
  \hspace*{\fill} 
  \begin{subfigure}{0.31\textwidth}
    \includegraphics[trim=1mm 0mm .5cm 0mm,width=\linewidth]{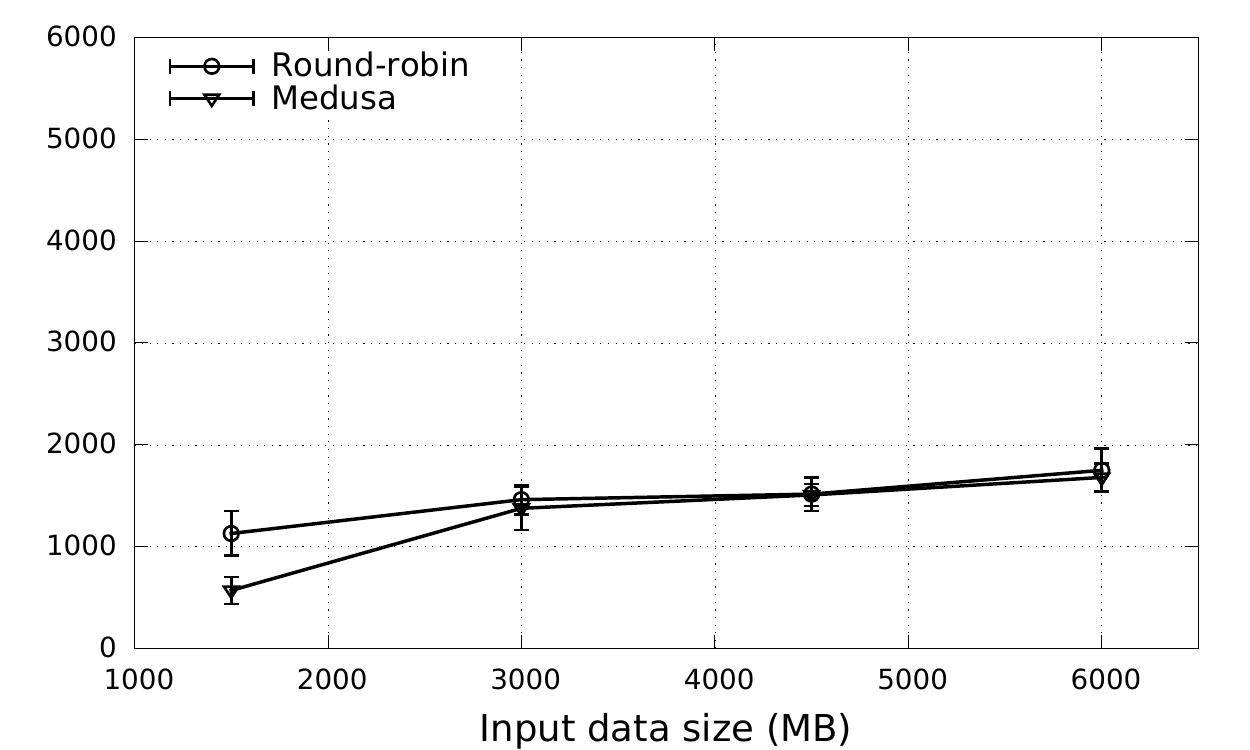}
    \caption{MonsterQuery}
    \label{fig:makespan:performance:trendline:monsterquery}
  \end{subfigure}
  \vspace{.5em}
  \caption{Job makespan of WordCount, WebdataScan, and MonsterQuery executions (no faults).}
  \label{fig:makespan:performance:trendline}
\end{figure*}

\paragraph{Performance without faults}
We first evaluate the performance of Medusa when no fault occurs during the entire job execution.
We consider $f=1$ in our experiments, so each job will be executed twice, \textit{i.e.}, $f+1$ times.
This choice of $f=1$ is based on the observation that the faults we consider in this paper, despite potentially having devastating consequences, are assumed not to be frequent.

\textit{Job efficiency.} Fig.~\ref{fig:makespan:performance:trendline} compares the job makespan of the Medusa scheduler against the baseline with different job sizes.
We observe that Medusa outperforms Round-robin in all cases.
In particular, notice that for the WordCount experiment (Fig.~\ref{fig:makespan:performance:trendline:wordcount}) \emph{the Medusa scheduler is up to 3 times faster to complete the job when compared with Round-robin for the larger input sizes}.
Moreover, from the  bars depicting the first and third quartiles of the makespan in both figures, we conclude that the \emph{Medusa scheduler offers more stable performance} as the variance of the Round-robin scheduler is much more evident (in fact, the variance of the others is so low that the  bars are barely perceptible).

The performance gain of Medusa in the other two experiments exists but is not so pronounced because the throughput between clouds is similar.
However, in terms of stability Medusa still offers an advantage: the variation in performance of the Round-robin is perceptible in all plots, contrary to our scheme.

\begin{figure}[t!]
  \centering
  \begin{subfigure}{.5\textwidth}
    \includegraphics[trim={0mm 0mm 0mm 1cm}, width=\linewidth]{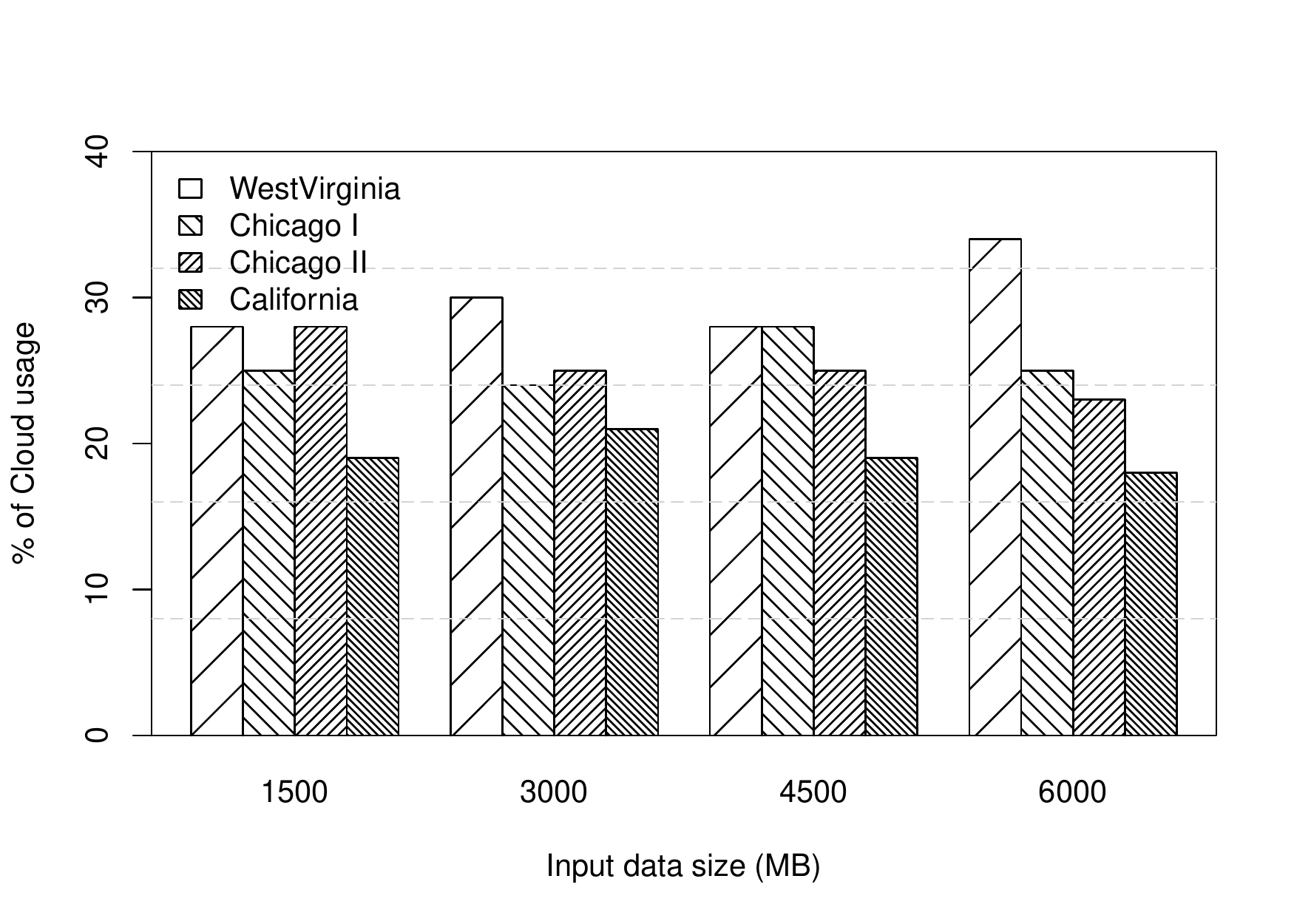}
    \vspace{-1.5\baselineskip}
    \caption{Round-robin}
    \label{fig:makespan:cloud:usage:rr}
  \end{subfigure}
  \vspace*{\fill} 
    \begin{subfigure}{.5\textwidth}
      \includegraphics[trim={0mm 0mm 0mm .5cm}, width=\linewidth]{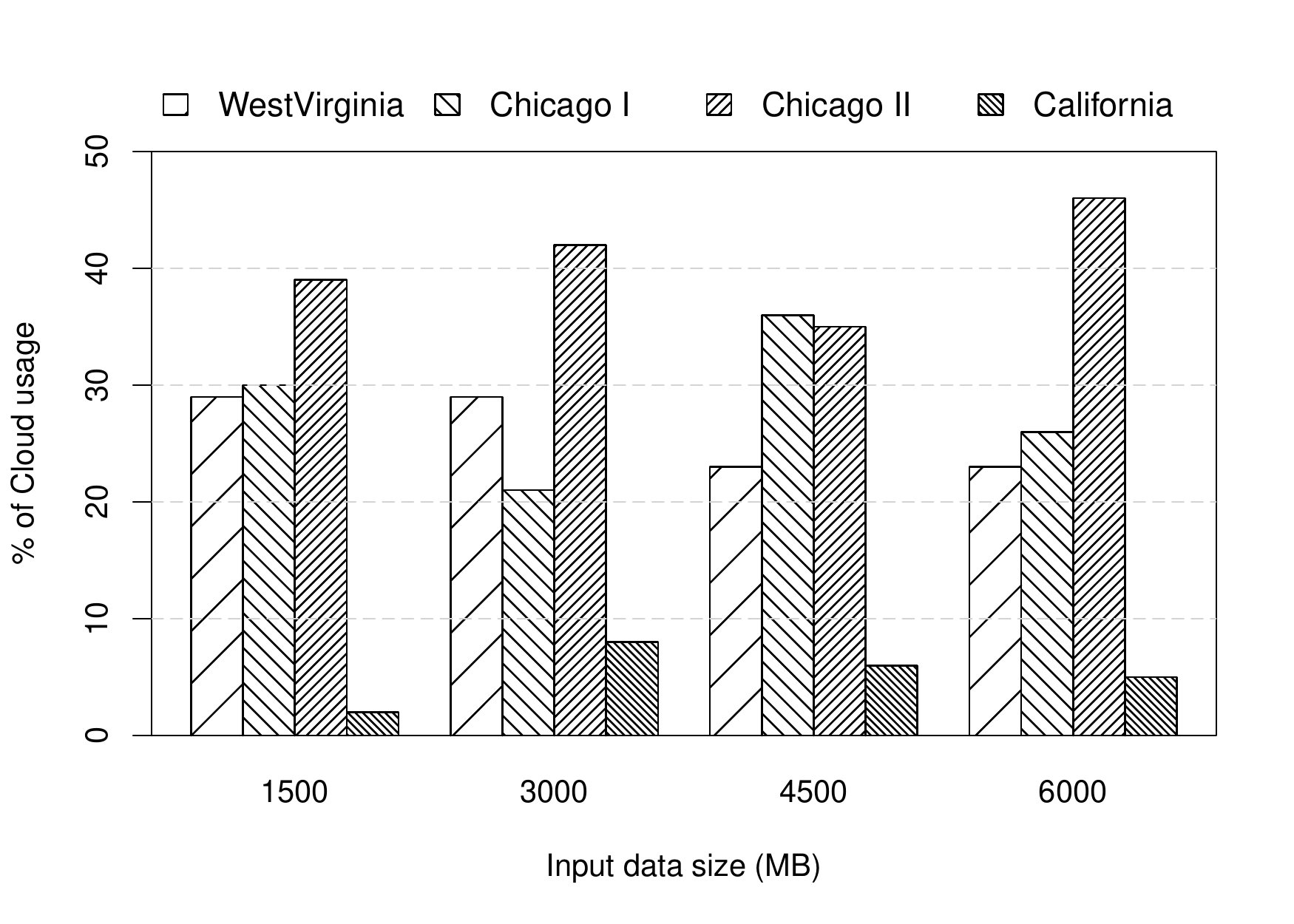}
    \vspace{-1.5\baselineskip}
    \caption{Medusa}
    \label{fig:makespan:cloud:usage:medusa}
  \end{subfigure}
  \vspace{.5em}
  \caption{Percentage of cloud usage.}
  \label{fig:cloud:usage}
\end{figure}
\begin{figure}
  \centering
  \includegraphics[trim=0mm 0mm 0mm 1cm,width=\linewidth]{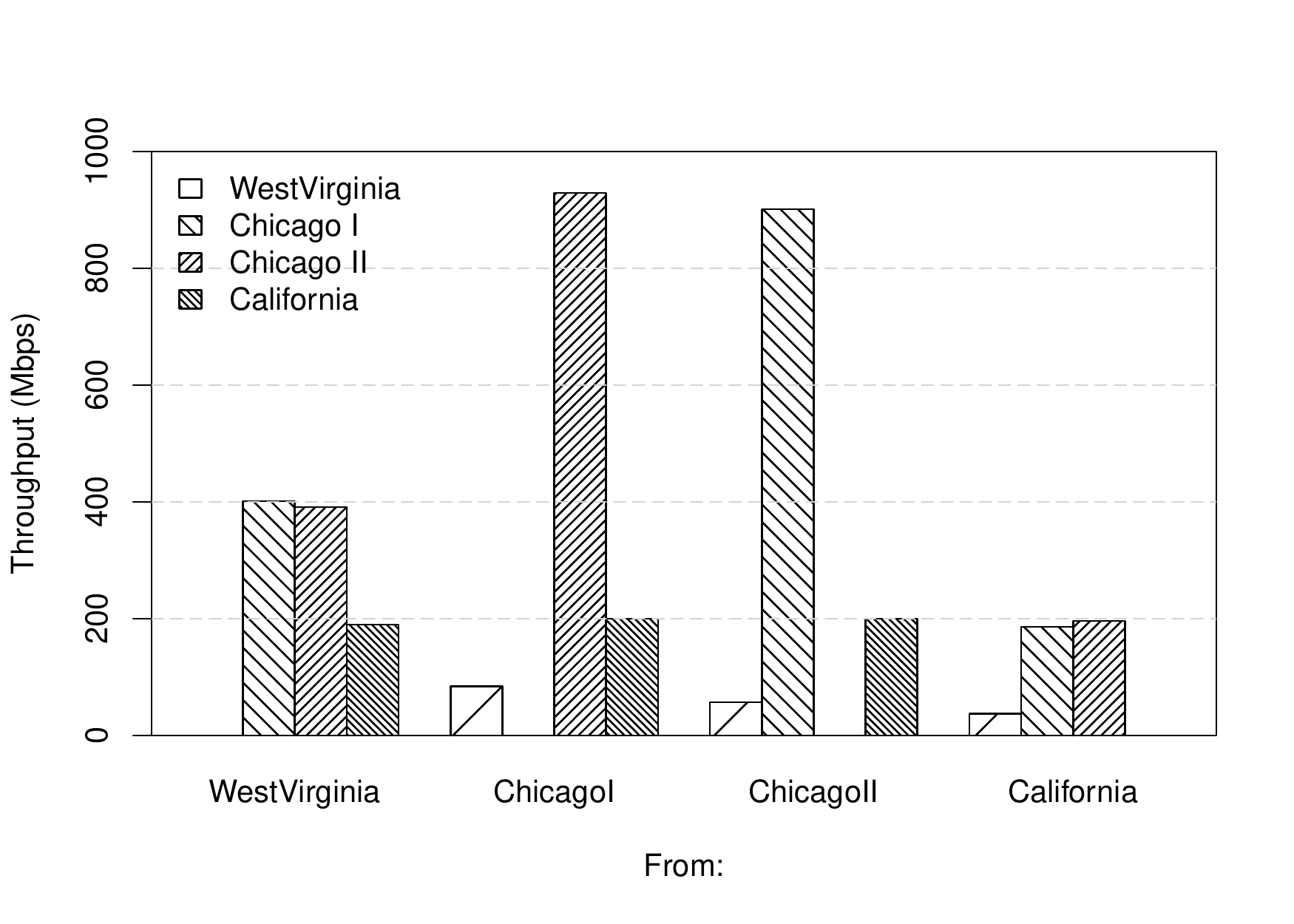}
  \caption{Throughput between each pair of clouds (WordCount).}
  \label{eval:fig:throughput:wordcount}
\end{figure}

\textit{Discussion.}
We now focus on trying to understand the results a bit further.
For this purpose we measured the load distribution among clouds with the WordCount application to understand the improved makespan of the Medusa scheduler in more detail.
Fig.~\ref{fig:cloud:usage} illustrates the usage of each cloud for executing an entire job when
different schedulers are used. The Round-robin scheduler distributes the workload evenly across the clouds, ignoring cloud and network performance.
In contrast, with the Medusa scheduler the jobs are launched more often in the Chicago clouds, followed by West-Virginia, and only rarely in California.
The reason is twofold.
First, as can be seen in Fig.~\ref{eval:fig:throughput:wordcount}, which shows the throughput measured between each pair of clouds in the system, the Chicago clouds have better network connections between them when compared to the other clouds.
Thus, the transfer time to these clouds is, on average, smaller, and this advantage results in the Chicago clouds being chosen more frequently to run replicated jobs.
We observe, however -- and that is the second reason --, that although West-Virginia has slightly lower network throughput when compared to California, more jobs are still scheduled to that cloud.
This is because the Medusa scheduler considers \emph{both} network transfer time and available computational capacity of the cloud when making its decision.
So, in this case, as the relative difference of network throughput is not very high, the cloud with higher computational power (West-Virginia in this case) is capable of offering a lower makespan of the entire job.

\begin{figure}
  \centering
  \includegraphics[trim=0mm 0mm 0mm 10mm,width=\linewidth]{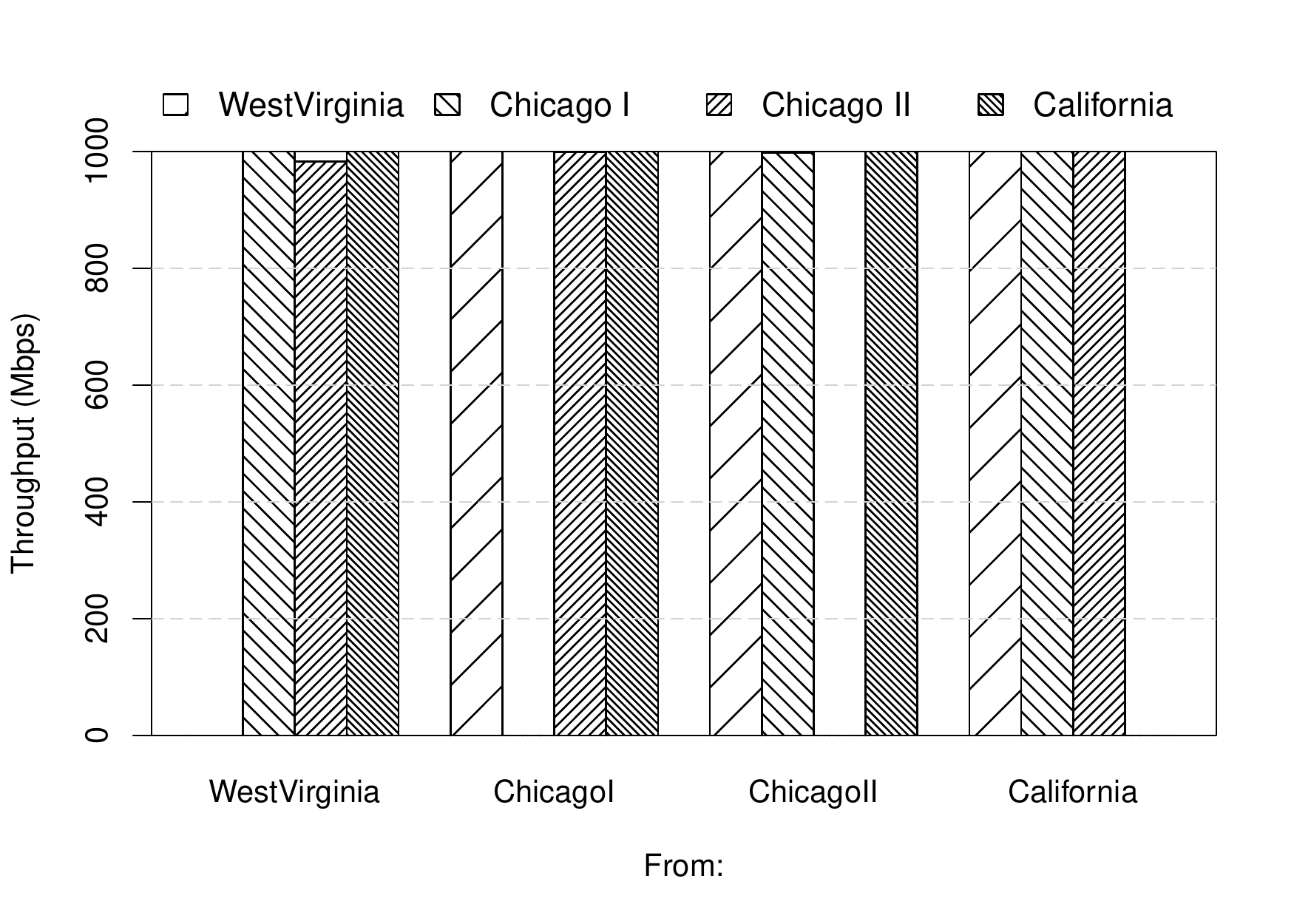}
  \caption{Throughput between each pair of clouds (WebdataScan and Monsterquery).}
  \label{eval:fig:throughput:others}
  \vspace{1em}
\end{figure}

Why does Medusa outperforms Round-robin so considerably in the WordCount experiment, in particular, when compared with the other two?
Recall from Section~\ref{subset:setup} that we ran the two sets of experiments in different sets of clouds.
Contrary to Fig.~\ref{eval:fig:throughput:wordcount}, the throughput measured between each pair of clouds in the WebdataScan and Monsterquery (Figure~\ref{eval:fig:throughput:others}) experiments was very stable.
The average throughput was around 1 Gbps.

As is clear, this scenario is more homogeneous in terms of network throughput between clouds, so a Round-robin scheduler performs reasonably well in this setting.
The reason why Medusa shows a very significant improvement for the WordCount experiment and less so for the others is therefore related to the heterogeneity of the scenario.

In summary, as we have shown, \textit{the combination of the cloud characteristics and network throughput both influence the scheduling decision explaining the best performance of Medusa over a traditional scheduler.}

\paragraph{Performance with faults}

In this section, we evaluate the performance of the Medusa scheduler when there are faults.
We consider $f=1$ in this experiment, as before.

\begin{figure}
  \centering
  \includegraphics[trim=5mm -2mm 5mm 0mm,width=.45\textwidth]{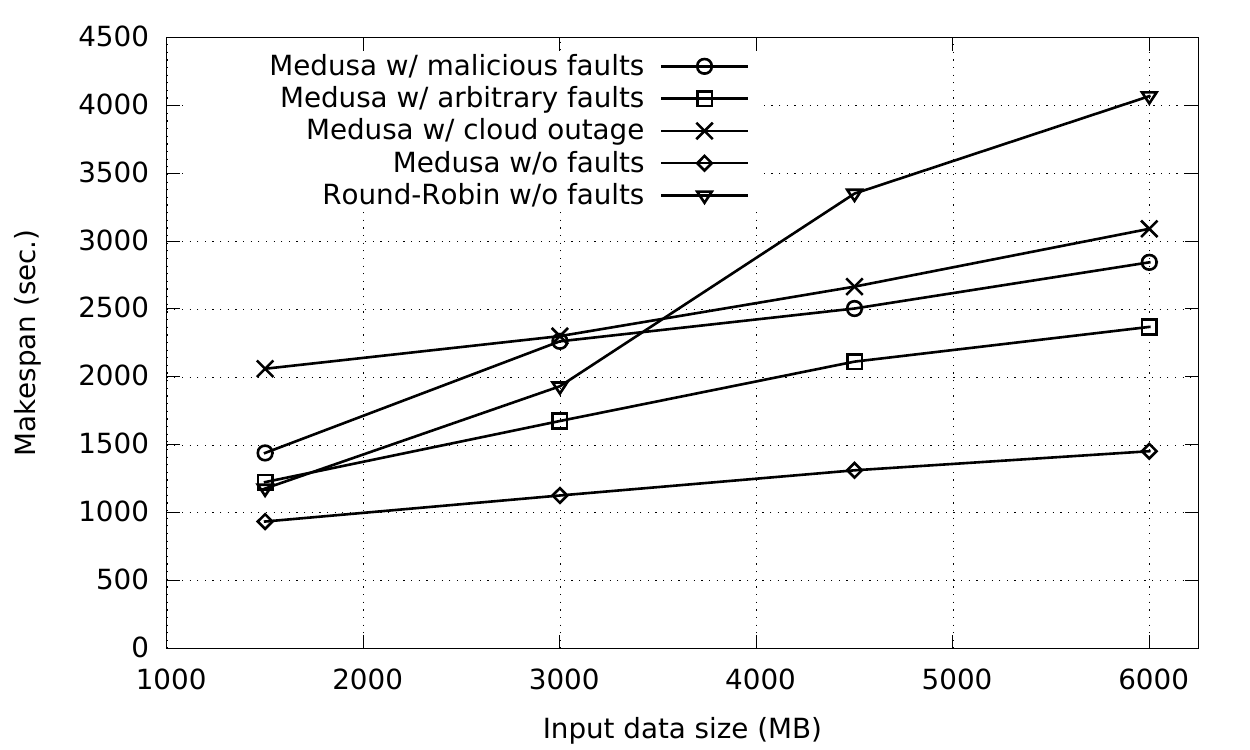}
  \caption{Job makespan with one fault injected (WordCount).}
  \label{eval:fig:performance:fault}
\end{figure}

We first evaluate the performance of the Medusa scheduler when a malicious fault occurs.
To this end, we inject a fault that corrupts the digest of a job output at the end of the vanilla WordCount execution, forcing the scheduler to launch an extra replica of this job in a different cloud.
Fig.~\ref{eval:fig:performance:fault} shows the performance of the Medusa scheduler when one such fault occurs (Medusa w/ malicious faults).
In this case, the job makespan doubles when compared to the case with no faults (Medusa w/o fault) as one extra job has to be executed in a different cloud.
This is due to the fact that the scheduler can not consider as options the cloud where the job has just run.
As explained in Section \ref{subsec:solution:scheduler}, that cloud cannot be trusted again.

We also evaluate the performance of the Medusa scheduler by assuming that the fault is arbitrary but not malicious (Medusa w/ arbitrary faults).
In this scenario, we also inject a fault that tampers the digest of one job output at the end of the vanilla WordCount execution.
The difference, in this case, is that the Medusa scheduler has the possibility to launch the extra replica of this job in the same cloud where the fault occurred (as we assume it is not malicious).
As we observe from Fig.\ref{eval:fig:performance:fault}, only tolerating arbitrary faults allows reducing the job makespan.
This is mainly because the extra replica of the faulty job will with high probability be scheduled to the same cloud, so there is no need to copy the input of this job to another cloud.

Finally, we evaluate the performance of the Medusa scheduler when one cloud outage happens (Medusa w/ cloud outage).
To this end, we simulate a cloud outage by crashing the resource manager of a random cloud.
This forces the scheduler to launch an extra replica of this job to a different cloud.
As shown in Fig.~\ref{eval:fig:performance:fault}, tolerating a cloud outage requires higher job makespan than tolerating arbitrary or malicious faults.
This is due to the scheduler taking some time to detect that the cloud has failed, looking for a copy of the data in another cloud, making the necessary data transfers and running the job.

Interestingly, when the input data is large, \emph{the Medusa scheduler in the presence of faults performs better than Round-robin with no faults occurring}.
We find it therefore unnecessary to report the performance of the Round-robin scheduler when a fault occurs.

To summarize, our experimental evaluation shows that the Medusa scheduler is more efficient than the conventional round-robin alternative, and achieves a significant gain in realistic scenarios of high cloud heterogeneity.

\section{Related Work}
\label{sec:relatedwork}

There has been much research on MapReduce since this framework was originally proposed in 2004~\cite{Dean:04}.
Just to give a few examples, there has been significant work on running MapReduce efficiently in different environments: multi-core/multiprocessor systems~\cite{Ranger:07}, heterogeneous environments~\cite{Zaharia:08}, mixed CPU-GPU environments \cite{conf-ipps-JiangA12}, high-latency eventually-consistent environments~\cite{Gunarathne:10}, among others.
As the related work on this subject is vast, we invite the reader for the survey by Lee et al.~\cite{Lee2012}.

With the requirements for compute-intensive analysis growing significantly, the need to scale-out MapReduce computation across clouds has lately given rise to interesting research on the subject.
G-Hadoop~\cite{Wang2013739} and G-MR~\cite{JayalathSE14} are two frameworks recently proposed to enable large-scale MapReduce computations to be distributed across clusters.
Our work shares the goal of scaling out to multiple clouds but has several differentiating points.
First, and most important, neither G-Hadoop nor G-MR deal with the types of faults our system tolerates: arbitrary and malicious faults, and cloud outages.
They deal solely with crash faults, similar to the original MapReduce.
In addition, they either require significant changes to Hadoop (G-Hadoop) or are a completely new Hadoop-based platform (G-MR).
In contrast, our proposal uses Hadoop and does not require any changes to this widely used platform.

These works show the importance of MapReduce and Hadoop.
Yet, neither advances the original platform from the fault tolerance point of view, as we propose in this paper.
We leverage the vast literature on dependability in distributed systems that appeared since the algorithms to tolerate Byzantine or arbitrary faults were introduced, 30 years ago~\cite{Lam82}.
State machine replication, for instance, is a generic solution to build systems that are crash or Byzantine fault-tolerant~\cite{Sch90}.
The practicality of implementing efficient Byzantine fault-tolerant replication was demonstrated in~\cite{Cas02}, and from then onwards other efficient algorithms have been proposed~\cite{Cle09b,Veronese:13,Ami06}.

We have proposed in previous work a Byzantine fault-tolerant MapReduce~\cite{Costa2013}.
Contrary to the system we propose in the current paper, the target of~\cite{Costa2013} was a single datacenter.
Hence, the system did not tolerate cloud outages or malicious faults caused by malicious insiders.
Notably, our previous solution required changes to Hadoop, contrary to the proposal we make here.
ClusterBFT is also a system for Byzantine fault-tolerant data-flow processing in clouds~\cite{conf/middleware/StephenE13}.
Contrary to our work, it was not designed for MapReduce.

\section{Concluding remarks} \label{sec:conclusion}

In this paper we proposed a platform, Medusa, for scaling out MapReduce to multiple clouds and, simultaneously, tolerating several kind of faults introduced by such multi-cloud environment.

Our proposal fulfilled the objectives we set forth.
First, Medusa scales MapReduce computations to multiple clouds.
Second, it extends the fault-tolerance offered by MapReduce to tolerate arbitrary and malicious faults, as well as cloud outages.
Third, it does so transparently to the user.
The Hadoop API is not touched and the existing Hadoop MapReduce programs run without modification.
Forth, Medusa is a proxy in the client, and thus the system does not require any modification to the Hadoop framework. The clouds just need to run ``vanilla'' Hadoop (\textit{e.g.}, Amazon Elastic MapReduce~\cite{AmazonEMR2015}).
Finally, as demonstrated by our extensive experimental evaluation in the ExoGENI testbed, it achieves this increased level of fault tolerance at a reasonable cost when compared with common alternatives. It does so by minimizing replication and by running a novel scheduling algorithm that judiciously chooses the best clouds to perform the necessary replicated jobs.

We made our code available as open source\footnote{https://bitbucket.org/pcosta\_pt/medusa\_hadoop}.
We invite the community to use it and improve upon our code.
As future work, we plan to investigate techniques to improve the performance of our system by replicating at a more fine-grained level than at a job level.



\bibliographystyle{ieeetr}
{\small
\setstretch{.95}
\bibliography{medusa-efficient-cloud-fault-tolerant-mapreduce-ieee}{}
}

\end{document}